\documentclass[final,3p,times,twocolumn]{elsarticle}
\biboptions{comma,sort&compress}
\usepackage{here}
\usepackage{graphicx}
\usepackage{epsfig}
\def\beq{\begin{equation}}
\def\eeq{\end{equation}}
\def\bea{\begin{eqnarray}}
\def\eea{\end{eqnarray}}

\journal{Nuc. Phys. (Proc. Suppl.)}

\begin{document}

\begin{frontmatter}

\title{Properties of the ground-state baryons in chiral perturbation theory}

\author[label1]{J. Martin Camalich\corref{cor1}}
\cortext[cor1]{Speaker}
\ead{camalich@ific.uv.es}
\author[label2,label3]{L.S. Geng}
\ead{lisheng.geng@ph.tum.de}
\author[label1]{J.M. Vicente Vacas}
\ead{vicente@ific.uv.es}
\address[label1]{Departamento de F\'{\i}sica Te\'orica and IFIC, Universidad de
Valencia-CSIC, Spain}
\address[label2]{School of Physics and Nuclear Energy Engineering, Beihang University, Beijing 100191, China}
\address[label3]{Physik Department, Technische Universit\" at M\"unchen, D-85747 Garching, Germany}

%\author{}

%\address{}

\begin{abstract}
\noindent
We review recent progress in the understanding of low-energy baryon structure by means of chiral perturbation theory. In particular, we discuss the application of this formalism to the description of various properties such as the baryon-octet magnetic moments, the electromagnetic structure of decuplet resonances and the hyperon vector coupling $f_1(0)$. Moreover, we present the results on the chiral extrapolation of recent lattice QCD results on the lowest-lying baryon masses and we predict the corresponding baryonic sigma-terms.
\end{abstract}


\end{frontmatter}
\section{Introduction}

Baryons are physical objects of great interest. Their properties and interactions are essential to understand those of the atomic nuclei or of more exotic kinds of systems like the strange matter, which is believed to play a role in the macroscopic properties of astrophysical objects, e.g. neutron stars~\cite{Weber:2006da}. On the other hand, baryon phenomenology allows to study the non-perturbative regime of Quantum Chromodynamics (QCD). It is a great scientific endeavor to understand the extremely rich spectroscopy and structure of baryons directly from the few parameters of QCD, namely the strong coupling constant and quark masses. Additionally, their weak decays and reactions provide information on the flavor structure of the electroweak interactions that eventually may point out departures from the Standard Model (SM) predictions in baryonic observables. 

Experiments on baryon spectroscopy, structure, decays or reactions are currently taking data or are planned in laboratories like CERN-SPS, GSI, etc,  and new facilities will be soon available at J-PARC, TJNAF,  LNF, etc. Moreover, the last few years have witnessed an impressive development in the Lattice QCD (LQCD) description of several observables and realistic results on baryon structure are starting to appear~\cite{Jansen:2008vs}. On the other hand the investigation of baryon phenomenology by means of the low-energy effective field theory of QCD, namely chiral perturbation theory ($\chi$PT) ~\cite{Gasser:1984gg,Gasser:1987rb}, has been troubled for a very long time because of several conceptual and technical problems like the poor convergence of the perturbative (chiral) series or the effects of the lowest-lying decuplet resonances. 

Recently, we have shown that a fairly good convergence is possible using a Lorentz covariant approach within a renormalization prescription of the loop-divergencies which recovers the power counting~\cite{Fuchs:2003qc} and is consistent with analyticity~\cite{Pascalutsa:2004ga}. We have systematically incorporated the decuplet resonances taking care of the power-counting and $consistency$ problems~\cite{Pascalutsa:2000kd,Geng:2009hh}. A model-independent understanding of diferent properties including the magnetic moments of the baryon-octet~\cite{Geng:2008mf,Geng:2009hh}, the electromagnetic structure of the decuplet resonances~\cite{Geng:2009ys} and the hyperon vector coupling $f_1(0)$~\cite{Geng:2009ik}, has been successfully achieved. Moreover, the quark mass dependence of LQCD results on the baryon masses has been studied in this approach~\cite{MartinCamalich:2010fp}, providing an extrapolation to the physical point that is better than using a linear extrapolation. We will briefly review these developments stressing the role they may play in the future. 

\section{Magnetic Moments}
\label{MMs}

The magnetic moments of the baryons are of the utmost importance since they contain information on the internal structure of baryons as read out by electromagnetic probes. A starting point is the $SU(3)_F$-symmetric model of Coleman and Glashow (CG)~\cite{Coleman:1961jn} that describes the baryon-octet magnetic moments in terms of two parameters. The success of this model relies on the almost preserved global $SU(3)_V$-symmetry of QCD with $u$, $d$ and $s$ flavors. The description of the symmetry-breaking corrections to the baryon magnetic moments can be addressed in a systematic and model-independent fashion by means of $\chi$PT. In this approach, the CG result appears naturally at leading-order (LO) as tree-level. At next-to-leading (NLO) order, there are only loop-contributions that depend on known couplings and masses and, therefore, no new undetermined low-energy constants (LECs) besides those appearing in the CG approach are to be included. The question is then whether the $SU(3)_F$-breaking corrections to the baryon-octet magnetic moments can be successfully addressed from a first principles approach by means of $\chi$PT; namely whether or not the chiral loops improve the classical CG results. A positive answer to this question has been given only recently~\cite{Geng:2008mf,Geng:2009hh} when applying the extended-on-mass-shell (EOMS) renormalization scheme ~\cite{Fuchs:2003qc}
which is an extension of $\overline{MS}$ developed to overcome the power-counting problem in the baryon sector of $\chi$PT. For a detailed presentation of our results and their comparison with heavy baryon (HB)~\cite{Jenkins:1990jv,Jenkins:1992pi}
or infrared (IR)~\cite{Becher:1999he,Meissner:1997hn} formulations of baryon $\chi$PT, see Refs.~\cite{Geng:2008mf,Geng:2009hh}. The comparison of our results with those obtained before stresses the importance in SU(3)$_F$-$\chi$PT of the relativistic corrections and of keeping unaltered the analytic properties of the theory. Concerning the inclusion of the decuplet resonances, $natural$ contributions that do not spoil the improvement over CG were found only in the EOMS framework and when the unphysical degrees of freedom contained in the relativistic spin-3/2 vector-spinor were removed by means of the $consistent$ couplings~\cite{Pascalutsa:2000kd}. It is also noteworthy that we obtain a good convergence since the NLO contribution is, at most, about one half of the LO one, what is consistent with our $a\;priori$ expectation that the contribution is of the order of $\sim m_\eta/\Lambda_{\chi SB}$.  

The aforementioned covariant approach that includes both octet and decuplet contributions has also been applied to the description of the electromagnetic structure of the decuplet resonances~\cite{Geng:2009ys}. In particular, the magnetic dipole moments of the $\Delta^+$ and $\Delta^{++}$ are predicted using the well-measured one of the $\Omega^-$ to
fix the only LEC appearing up to NLO
\begin{equation}
  \mu_{\Delta^{++}}= 6.0(1.0)\mu_N,\hspace{1cm}\mu_{\Delta^+}=2.84(50)\mu_N, 
\end{equation}
where the error bars are an estimation of higher-order contributions obtained looking at the ratio between NLO and LO contributions (we take $50\%$ of the NLO over LO ratio)~\cite{Geng:2009ys}. 
The relevance of these results lies on the ongoing efforts from the experimental side to measure the magnetic moments of these two resonances~\cite{Machavariani:1999fr,Kotulla:2002cg,Kotulla:2008zz}. On the theoretical side, calculations from many different approaches have arisen in the last decades~\cite{Geng:2009ys}. Our results are compatible with the values quoted by the Particle Data Group~\cite{Amsler:2008zzb} and the agreement with the latest experimental analysis, $\mu_{\Delta^{++}}=6.14\pm0.51\mu_N$~\cite{LopezCastro:2000cv}, is excellent. Finally, the electromagnetic properties related with the higher-order multipoles in the expansion of the spin-3/2 electromagnetic vertex, namely the electric quadrupole moment and the magnetic octupole moment, have been also predicted~\cite{Geng:2009ys}.

\section{Hyperon vector coupling $f_1(0)$}

\begin{table*}
\centering
      \renewcommand{\arraystretch}{1.5}
     \setlength{\tabcolsep}{0.2cm}
\caption{Results on the relative $SU(3)_F$-breaking of $f_1(0)$ in $\%$ for different channels obtained in $\chi$PT up to NNLO including octet and decuplet
contributions and those obtained in other approaches.\label{Table}}
\begin{tabular}{c|cccccc|}
\cline{2-7}
&B$\chi$PT&HB$\chi$PT&Large $N_c$&QM&$\chi$QM&LQCD\\
\hline
\multicolumn{1}{|c|}{$\Lambda \;N$}&$+0.1^{+1.3}_{-1.0}$&+5.8&$+2\pm2$&$-1.3$&+0.1&\\
\multicolumn{1}{|c|}{$\Sigma\;N$}&$+8.7^{+4.2}_{-3.1}$&+9.3&$+4\pm3$&$-1.3$&+0.9&$-1.2\pm2.9\pm4$\\
\multicolumn{1}{|c|}{$\Xi\Lambda$}&$+4.0^{+2.8}_{-2.1}$&+8.4&$+4\pm3$&$-1.3$&+2.2&\\
\multicolumn{1}{|c|}{$\Xi\Sigma$}&$+1.7^{+2.2}_{-1.6}$&+2.6&$+8\pm5$&$-1.3$&+4.2&$-1.3\pm1.9$\\
\hline
\end{tabular}
\end{table*}

The Cabibbo-Kobayashi-Maskawa (CKM) matrix ~\cite{Cabibbo:1963yz,Kobayashi:1973fv} plays a very important role in our study and understanding of flavor physics. In particular, its low mass sector allows for a precise test of the Standard Model through the CKM unitarity relation,
\begin{equation}
|V_{ud}|^2+|V_{us}|^2+|V_{ub}|^2=1,\label{Unitarity}
\end{equation}
where one needs accurate values for $V_{ud}$, $V_{us}$, and $V_{ub}$. Among them, $V_{ub}$ is quite small and can be
neglected at the present precision. The element  $V_{ud}$ can be obtained from superallowed nuclear beta, neutron and pion decays, whereas $V_{us}$ can be extracted from kaon, hyperon, and tau decays (for a recent review, see Ref.~\cite{Amsler:2008zzb}). We now focus on how to determine $V_{us}$ from hyperon semileptonic decay data. 

The hyperon matrix elements of the weak flavor-changing currents are described by three vector (axial) form factors $f_i(q^2)$ ($g_i(q^2)$) with $i=1,2,3$. The decay ratio of the semileptonic decay $B\rightarrow b l\bar{\nu}$ will then be determined by these form factors, the Fermi constant $G_F$, and the CKM element $V_{us}$. Indeed, if we define as a relevant $SU(3)_F$-breaking parameter
$\beta=\frac{M_B-M_b}{M_B}$, we can perform a power expansion of the decay rate about the $SU(3)_F$-symmetric limit
\begin{eqnarray}
&R\sim G_F^2 V_{us}^2 \left(\left(1-\frac{3}{2}\beta+\frac{6}{7}\beta^2\right)f_1^2+\frac{4}{7}\beta^2f_2^2\right.\nonumber\\
&\left.+\left(3-\frac{9}{2}\beta+\frac{12}{7}\beta^2\right)g_1^2+\frac{12}{7}g_2^2+\right.\nonumber\\
&\left.\frac{6}{7}\beta^2f_1f_2+\left(-4\beta+6\beta^2\right)g_1g_2 +\mathcal{O}(\beta^3)\right)\label{Eq:Rat3}
\end{eqnarray}
where the form factors are evaluated at $q^2=0$, although a linear $q^2$ dependence in $f_1$ and $g_1$
must also be considered at this order~\cite{FloresMendieta:2004sk}. Moreover, the $SU(3)_F$-symmetric limit for 
$f_2$ can be used. The most relevant contributions to the ratio come then from $g_1$, $f_1$ and also $g_2$. Therefore, in order to extract accurately $V_{us}$ from semileptonic hyperon decay data, one requires to understand, in a model-independent fashion, the SU(3)-breaking contributions to these moments. The $g_2$ vanishes in the $SU(3)_F$-symmetric limit, and we will assume $g_2=0$. The axial charge $g_1$, which is described in the symmetric limit by the parameters $D$ and $F$, receives $\mathcal{O}(\beta)$ breaking corrections. Nevertheless, as it has been proposed in Ref.~\cite{Cabibbo:2003cu}, we can use the measured $g_1/f_1$ ratios as the basic experimental data to equate $g_1$ in terms of $f_1$ in Eq.~(\ref{Eq:Rat3}). On the other hand, $f_1$ is protected by the Ademollo-Gatto Theorem~\cite{Ademollo:1964sr} which states that  breaking corrections start at $\mathcal{O}(\beta^2)$. 

The Ademollo-Gatto theorem is a consequence of the underlying $SU(3)_V$ symmetry of QCD, which has also important consequences when addressing a calculation of $f_1(0)$ in $\chi$PT. Namely, one finds that no unknown LECs contributing to this vector charge are allowed until chiral order $\mathcal{O}(p^5)$. Therefore, a loop calculation up to and including NNLO
only depends on known masses and couplings and is a genuine prediction of $\chi$PT. Moreover, there are not divergencies 
or power counting breaking terms up to this order so that a counting restoration procedure does not seem necessary in this case. This program has been developed in different steps along the last two decades~\cite{Anderson:1993as,Kaiser:2001yc,Krause:1990xc,Lacour:2007wm,Villadoro:2006nj,Geng:2009ik}. A full NNLO calculation including both octet and decuplet contributions in the covariant framework has been undertaken recently~\cite{Geng:2009ik}. In the latter work the problem with the convergence found in the HB calculation of Ref.~\cite{Villadoro:2006nj} has also been explained and fixed. 

In Table~\ref{Table} we present the results for the relative $SU(3)_F$-breaking correction $100\left(\frac{f_1(0)}{f_1^{SU(3)}(0)}-1\right)$ in covariant $\chi$PT (B$\chi$PT) and HB$\chi$PT including octet and decuplet contributions up to NNLO. We also present those obtained in Large $N_c$~\cite{FloresMendieta:2004sk}, in a quark model (QM)~\cite{Donoghue:1986th}, in a chiral quark model ($\chi$QM)~\cite{Faessler:2008ix} and in LQCD
~\cite{Guadagnoli:2006gj,Sasaki:2008ha}. The error bars in the B$\chi$PT are an estimation of higher order uncertainties~\cite{Geng:2009ik}.
The results quoted from  Ref.~\cite{Donoghue:1986th} are quite general in quark model calculations and reflect the 
naive expectation that  $SU(3)_F$-breaking corrections, at least for the $\Sigma N$ channel, should be negative. On the other hand, the different chiral approaches agree in the positive sign and the approximate size of these corrections, what may indicate the non-triviality of the multiquark effects induced by the chiral dynamics. It is also remarkable the agreement with those obtained in a different systematic approach to non-perturbative QCD as the Large $N_c$. The results of lattice QCD are marginally compatible with ours although they favor negative corrections to $f_1(0)$. However, it must be pointed out that the pion masses in these simulations are still rather high, namely $\sim$400 MeV for $\Xi^0\rightarrow\Sigma^+$~\cite{Sasaki:2008ha} and $\sim$700 MeV for $\Sigma^-\rightarrow
n$~\cite{Guadagnoli:2006gj}. Another issue to be highlighted is the chiral extrapolation of the lattice QCD results to the physical point, for which our results might be helpful in the future. And the other way around, the lattice QCD could provide information about the higher-order local contributions in the chiral approach and could reduce the theoretical uncertainty of the B$\chi$PT calculation~\cite{Guadagnoli:2006gj}. In any case, a lattice simulation close to the physical point will be very helpful and eventually conclusive about the nature of the $SU(3)_F$-breaking corrections to $f_1(0)$. 

With the elements developed above we obtain a determination of the CKM element $V_{us}$  that $combines$ the information on 
the different channels and includes the experimental errors~\cite{Mateu:2005wi} and higher-order errors estimated for $f_1(0)$ in B$\chi$PT
~\cite{Geng:2009ik}
\begin{equation}
V_{us}=0.2176\pm0.0029\pm\Delta_V, \label{vus}
\end{equation}
where $\Delta_V$ accounts for other systematic uncertainties. At the order we work in Eq.~(\ref{Eq:Rat3}), this uncertainty is due to the $SU(3)_F$-breaking correction to $g_2$ that has not been considered. This contribution is $\sim\mathcal{O}(\beta^2)$  and potentially as important for the extraction of $V_{us}$ as the $SU(3)_F$-breaking correction to $f_1$. 

We first compare our result with other determinations obtained from the decay rates and $g_1/f_1$ in the hyperon semileptonic data; namely,  $V_{us}=0.2199\pm0.0026$ in Large $N_c$~\cite{FloresMendieta:2004sk}  and $V_{us}=0.2250(27)$ in the $SU(3)_F$-symmetric model~\cite{Cabibbo:2003cu}. The comparison with the latter indicates the sensitivity to a breaking correction to $f_1(0)$ of $\sim\mathcal{O}(\beta^2)$ and suggests that the $SU(3)_F$-symmetric assumption is not reliable enough for the accuracy required by the determination of $V_{us}$. The agreement between the B$\chi$PT and the Large $N_c$ is a consequence of the consistency shown in Table~\ref{Table} and of the fact that in both approaches the $SU(3)_F$-breaking correction
to $g_2$ have been ignored.

On the other hand, our result is somewhat smaller than the ones obtained from kaon and tau decays or from the $f_K/f_\pi$ ratio~\cite{Amsler:2008zzb}. It is not compatible either with the unitarity condition Eq. (\ref{Unitarity}) when using the value obtained from superallowed beta decays~\cite{Amsler:2008zzb}. Nonetheless, the result shown in Eq. (~\ref{vus}) is not complete and has to be improved with the model-independent description of the $SU(3)_F$-breaking corrections to $g_2$. As argued in Ref.~\cite{Cabibbo:2003cu}, the trends shown by $\Sigma^-\rightarrow n$ and $\Lambda\rightarrow p$ data indicate that the incorporation of the $SU(3)_F$-breaking corrections to $g_2$ will raise the value of $V_{us}$ in these two channels.  Unfortunately, the data for hyperon decays is not yet precise enough to address a quantitative study of this form factor. From the theoretical side, a determination of these corrections in lattice QCD and an analysis in B$\chi$PT would be useful to ascertain the effects that $g_2$ may have on the determination of $V_{us}$.

\section{Baryon masses}

\begin{figure}[t]
\centering
\includegraphics[width=\columnwidth]{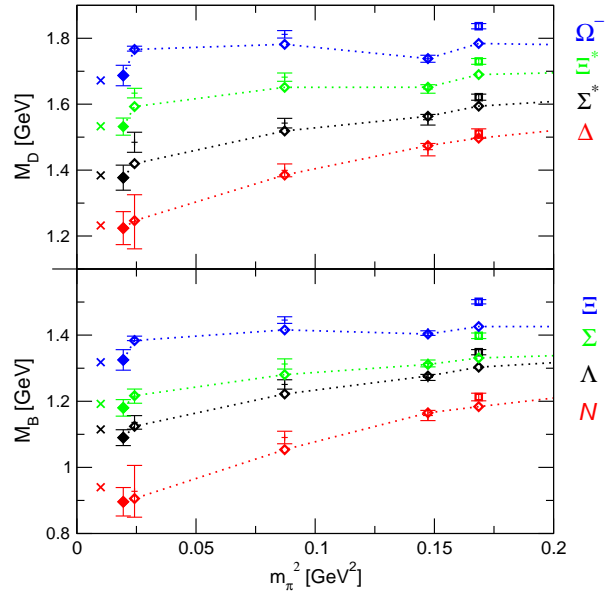}
\caption{Extrapolation of the PACS-CS results~\cite{Aoki:2008sm} on the lowest-lying baryon masses within the 
covariant formulation of $SU(3)_F$-B$\chi$PT up to NLO. The LQCD points used in the 
fit are represented with the corresponding error bars which do not include the correlated uncertainties. The lattice points in $m_\pi^2\simeq0.15$ GeV$^2$ involve a lighter strange quark mass. The diamonds denote our  
results after the fit and they are connected by a dotted line added to guide the eye.
The boxes are lattice points not included in the fit (heavier kaon mass) and the filled diamonds are the extrapolated values which are to be compared with experimental data (crosses). The latter are slightly shifted for a better comparison with the extrapolation results. \label{fig_graph}}
\end{figure}

In the last decades there has been a sustained interest in the description of the lowest-lying baryon mass spectrum by means of $SU(3)_F$-B$\chi$PT (see e.g. Refs.~\cite{Jenkins:1991ts,Bernard:1993nj,Borasoy:1996bx,WalkerLoud:2004hf,Tiburzi:2004rh,Lehnhart:2004vi,MartinCamalich:2010fp}). The chiral corrections to the Gell-Mann-Okubo baryon octet mass relation and Gell-Mann's decuplet equal spacing rules (we denote generically as GMO),
\begin{eqnarray}
&&\hspace{0.9cm}3M_\Lambda+M_\Sigma-2M_N-2M_\Xi=0\label{Eq:GMOBar}\\
&&M_{\Sigma^*}-M_\Lambda=M_{\Xi^*}-M_{\Sigma^*}=M_{\Omega^-}-M_{\Xi^-}\label{Eq:DES} 
\end{eqnarray}
have received special attention. More specifically, the GMO relations, which are recovered in $\chi$PT at LO, are known to work with an accuracy of $\sim7$ MeV. A puzzling and not yet well understood feature of the leading chiral corrections is that they preserve the GMO equations within $\sim10$ MeV whereas the corrections to any of the individual baryon masses are of order $\sim$100-1000 MeV~\cite{Jenkins:1991ts}. Another interesting aspect is that the analysis of the baryon masses provides hints on their scalar structure, i.e. the sigma terms~\cite{Pagels:1974se,Bernard:1993nj}. These magnitudes, besides providing a measure of the explicit symmetry breaking and of the meson-cloud contribution to the baryon masses~\cite{Pagels:1974se}, are relevant for other areas of phenomenology~\cite{Ellis:2008hf,Giedt:2009mr}. 

\begin{table*}[t]
\renewcommand{\arraystretch}{1.3}     
\setlength{\tabcolsep}{0.1cm}
\centering
\caption{Values of the LECs in the baryon-octet sector from the fits to the experimental, the PACS-CS~\cite{Aoki:2008sm} and the LHP~\cite{WalkerLoud:2008bp} results on the baryon masses using Lorentz covariant B$\chi$PT up to NLO. \label{Table:LECsOct}}
\begin{tabular}{c|ccc|cc|}
\cline{2-6}
&$M_{B0}$ [GeV]&$b_0$ [GeV$^{-1}$]&$M_{B0}^{eff}$ [GeV]&$b_D$ [GeV$^{-1}$]&$b_F$ [GeV$^{-1}$]\\
\hline
\multicolumn{1}{|c|}{Expt.}&-&-&1.840(5)&0.199(4)&$-$0.530(2)\\
\multicolumn{1}{|c|}{PACS-CS}&0.756(32)&$-0.978(38)$&1.76(7)&0.190(24)&$-$0.519(19)\\
\multicolumn{1}{|c|}{LHP}&0.780(31)&$-1.044(45)$&1.85(8)&0.236(24)&$-$0.523(21)\\
\hline
\end{tabular}
\end{table*}

\begin{table*}[t]
\renewcommand{\arraystretch}{1.3}     
\setlength{\tabcolsep}{0.1cm}
\centering
\caption{Values of the LECs in the baryon-decuplet sector from the fits to the experimental, the PACS-CS~\cite{Aoki:2008sm} and the LHP~\cite{WalkerLoud:2008bp} results on the baryon masses using Lorentz covariant B$\chi$PT up to NLO. \label{Table:LECsDec}}
\begin{tabular}{c|ccc|c|}
\cline{2-5}
&$M_{T0}$ [GeV]&$t_0$ [GeV$^{-1}$]&$M_{T0}^{eff}$ [GeV]&$t_D$ [GeV$^{-1}$]\\
\hline
\multicolumn{1}{|c|}{Expt.}&-&-&1.519(2)&$-$0.694(2)\\
\multicolumn{1}{|c|}{PACS-CS}&954(37)&$-1.05(8)$&1.49(8)&$-$0.682(20)\\
\multicolumn{1}{|c|}{LHP}&944(42)&$-1.28(8)$&1.60(8)&$-$0.609(14)\\
\hline
\end{tabular}
\end{table*}

On the other hand, LQCD calculations of the lowest-lying baryon mass spectrum have been undertaken by different collaborations using $N_f=2+1$ dynamical actions with light quark masses close to the physical point~\cite{WalkerLoud:2008bp,Aoki:2008sm,Lin:2008pr,Durr:2008zz,Bietenholz:2010jr}. The LHP~\cite{WalkerLoud:2008bp} and PACS-CS~\cite{Aoki:2008sm} collaborations have reported tremendous difficulties to understand the quark mass dependence and the chiral extrapolation of their results within HB$\chi$PT. This problem has been recently revisited in Ref.~\cite{MartinCamalich:2010fp}. In sharp contrast with the results obtained using the heavy-baryon expansion, it has been found that a good description of the LQCD results can be achieved within the Lorentz covariant approach to $SU(3)_F$-B$\chi$PT up to NLO. Moreover, the values of the masses extrapolated to the physical point of quark masses are manifestly better at NLO than those obtained using the linear extrapolation given by the GMO approach at LO. The study of the results of the LHP collaboration~\cite{WalkerLoud:2008bp} confirm all these conclusions. 

In Fig.~\ref{fig_graph}, we show the quark mass dependence and extrapolation of the lowest-lying baryon masses in Lorentz covariant $\chi$PT for the case of the analysis of the PACS-CS results~\cite{Aoki:2008sm}. The improvement obtained at NLO in covariant $SU(3)_F$-B$\chi$PT, highlights the effect of the leading chiral non-analytical terms in the extrapolation even from light quark masses as small as those used by PACS-CS.~\cite{Aoki:2008sm} ($m_\pi\simeq156$ MeV). On the other hand, the comparison between the results obtained in covariant and HB results~\cite{MartinCamalich:2010fp} illustrates the importance of the relativistic corrections in the understanding of the dependence shown by the lattice simulations on the baryon masses at relatively heavy quark-masses. This is specially true for the extrapolation of the LHP results, that are quite far away from the physical point of quark masses ($m_\pi\gtrsim293$ MeV). 

An important issue concerns the determination of the LECs of $SU(3)_F$-B$\chi$PT using 2+1-flavor 
simulations. In Tables~\ref{Table:LECsOct} and~\ref{Table:LECsDec}, we compare the values of the LECs determined studying the experimental values of the baryon masses with those obtained when fitting the corresponding quark-mass dependence of the PACS-CS or the LHP results. Since the experimental data do not disentangle $M_{B0}$ ($M_{T0}$) from $b_0$ ($t_0$) for the baryon-octet (-decuplet), in the comparison with the experimental determinations we must consider the effective masses $M_{B0}^{eff}$ ($M_{T0}^{eff}$) instead of these LECs~\cite{MartinCamalich:2010fp}.

In the case of the baryon-octet masses, the values of the LECs determined using either of the two LQCD sets of results partially agree with each other and they both are consistent with those resulting from the experimental determination. This suggests a non-trivial consistency in the baryon-octet sector between the lattice actions employed by the two collaborations (at different lattice spacings) and the experimental information on the masses through covariant B$\chi$PT up to NLO of accuracy. For the masses of the decuplet-baryons, while the values obtained using the PACS-CS results agree with those determined with the experimental data, the fit to the LHP results presents a value of $t_D$ that is not consistent with the experimental one. Some problems on these LQCD results for the decuplet masses were already noticed by the LHP collaboration~\cite{WalkerLoud:2008bp}.

\begin{table}[t]
\renewcommand{\arraystretch}{1.3}     
\setlength{\tabcolsep}{0.3cm}
\centering
\caption{Predictions on the $\sigma_{\pi N}$ and $\sigma_{sN}$ terms (in MeV) of the baryon-octet in covariant $SU(3)_F$-B$\chi$PT by fitting the LECs to the PACS-CS~\cite{Aoki:2008sm} or LHP~\cite{WalkerLoud:2008bp} results.   \label{Table:ResSigmasB}}
\begin{tabular}{c|cc|}
\cline{2-3}
&PACS-CS&LHP\\
\hline
\multicolumn{1}{|c|}{$\sigma_{\pi N}$}&59(2)(17)&61(2)(21)\\
\multicolumn{1}{|c|}{$\sigma_{sN}$}&$-7$(23)(25)&$-4$(20)(25)\\
\hline
\end{tabular}
\end{table}

A reliable combination of LQCD and $\chi$PT becomes a powerful framework to understand hadron phenomenology from first principles and may have sound applications. This can be illustrated in the scalar sector with the determination of the sigma terms from the analysis of the masses through the Hellman-Feynman theorem,
\begin{equation}
\sigma_{\pi \mathcal{B}}=m\frac{\partial M_\mathcal{B}}{\partial m}, \hspace{1cm}\sigma_{s \mathcal{B}}=m_s\frac{\partial M_\mathcal{B}}{\partial m_s}. \label{Eq:sigmasBH-F}
\end{equation}
In Table ~\ref{Table:ResSigmasB} we present the results on $\sigma_{\pi N}$ and $\sigma_{sN}$ after fitting the LECs to the PACS-CS and LHP results and with the uncertainties determined as has been discussed above for the masses. It is interesting to note that our results are in agreement with those of Ref.~\cite{Young:2009zb} obtained within the cut-off renormalized B$\chi$PT. 

\section{Conclusions}

In this work we have reported the recent progress that has been achieved in the model-independent description of diverse baryon phenomenology by means of $\chi$PT. More precisely, we have shown that a reasonable chiral convergence can be achieved using a Lorentz covariant approach to B$\chi$PT that systematically includes decuplet contributions. The electromagnetic structure of the baryons, the vector hyperon coupling $f_1(0)$ and the baryon masses comprise the properties studied so far. Besides giving a successful description of the experimental data concerned in these examples, the relativistic corrections have shown to be necessary in order to understand the quark mass dependence of the LQCD results on the baryon masses. Further applications of the Lorentz covariant approach in baryon phenomenology and in the analysis of LQCD simulations are foreseeable. Finally, it is worth noticing that this approach has been extended to heavy-light systems~\cite{Geng:2010vw}.

\section{Acknowledgments}

This work was partially supported by the  MEC contract  FIS2006-03438 and the EU Integrated Infrastructure Initiative Hadron Physics Project contract RII3-CT-2004-506078. JMC acknowledges the MICINN for support. LSG's work was partially supported by the Fundamental Research Funds for the Central Universities, the Alexander von Humboldt foundation, and the MICINN in the program Juan de la Cierva.



\end{document}